\documentstyle{qcdparis}
\input epsf
\begin{document}
\pagestyle{plain}
\title{QCD and Diffraction in DIS}
\author{Leonid Frankfurt$^{\dag}$ and Mark Strikman$^{\dagger}$}
\affil{$^{\dag}$ School of Physics and
  Astronomy Raymond and Beverly Sackler Faculty of Exact Sciences,\\Tel Aviv
University,
Israel\\
$^{\dagger}$  Physics Department, Pennsylvania State University
University Park, PA 16802, U.S.A.  }

\abstract{Coherence phenomena,  and the
non-universality of parton structure of the effective Pomeron are explained.
New hard phenomena directly calculable in QCD such as diffractive
electroproduction of states with $M^2\ll Q^2$  as well as new options to
measure the light-cone wave functions of
various hadrons are considered. An analogue of Bjorken scaling is predicted
for the diffractive electroproduction of $\rho$ mesons at large momentum
transfers and for the production of large rapidity gap events, as observed at
HERA.  A phenomenological QCD evolution equation is suggested to calculate the
basic characteristics of the large rapidity gap events. The increase of parton
densities at small $x$ as well as new means to disentangle experimentally soft
and hard physics are considered. We discuss constraints on the increase of
deep inelastic amplitudes with $Q^2$ derived from  unitarity of the S
matrix for collisions of wave packets. New ways to probe QCD physics of hard
processes at large longitudinal distances and to answer the long standing
problems on the origin of the Pomeron are suggested.  Unresolved problems and
perspectives of small $x$ physics are also outlined.}

\resume{ Nous presentons une revue sur le role respectif de QCD dure et
           molle dans les reactions diffractives inclusives et exclusives. }

\twocolumn[\maketitle]
\fnm{7}{Talk given in the
plenary  session
at the Workshop on Deep Inelastic scattering and QCD,
Paris, April 1995}

\section{Introduction}

The aim of this talk is to outline QCD predictions for color
coherence phenomena -- a result of nontrivial interplay of hard and
soft QCD physics specific for high energy processes (for more detailed
discussion see
\cite{AFS}. Coherence
phenomena provide an important link between the well understood
physics of hard processes and the physics of soft processes which at
present is mostly phenomenological.  The soft/hard interplay is
elaborated for the exclusive deep inelastic processes $\gamma^*_L + N
\rightarrow a +N$ for $M_a^2 \ll Q^2$ directly calculable in QCD.
These processes provide new methods of investigating the structure of
hadrons and the origin of the Pomeron and allow to search for new
forms of hadronic matter in heavy ion collisions (for a review and
references see~\cite{FMS94}). The phenomenon of coherence reveals
itself in high energy processes through a large probability of
occurrence of diffractive processes and through their specific
properties.  Thus in this report we concentrate mostly on diffractive
processes.

\section{ Interaction  cross section for small size wave packet.}

One of the striking QCD predictions for hard processes dominated by
large longitudinal distances is that if a hadron is found in a small
size configuration of partons it interacts with a target with a small
cross section .
The
prediction which follows from the factorization
theorem for hard processes in QCD is in variance with many
phenomenological approaches based on pre-QCD ideas and on quark models
of hadrons.

A sufficiently energetic wave packet with zero baryon and color
charges localized in a small transverse volume in the impact parameter
space can be described by a $q\bar q$ pair.  This conclusion follows
from asymptotic freedom in QCD which implies that the contribution of
other components is suppressed by a power of the strong coupling
constant $\alpha_s$ and/or a power of $Q^2$.  A familiar example of
such a wave packet is a highly virtual longitudinally polarized
$\gamma^*$ in a $q\bar q$ state.  Within the parton model the cross
section for the interaction of such a photon with a target is
suppressed by an additional  power of $Q^2$.  But at the same time the
probability
for a longitudinal photon to be in a large transverse size
configuration (soft physics=parton model contribution) is suppressed
by a power of $Q^2$.  These properties explain why reactions initiated
by longitudinally polarized photons are best to search for new QCD
phenomena.

The cross section for a high-energy interaction of a small size $q\bar
q$ configuration off any target can be unambiguously calculated in QCD
for low $x$ processes by applying the QCD factorization theorem.  In the
approximation when the leading
$\alpha_s \ln {Q^2\over \Lambda_{QCD}^2}
\ln x$ terms are accounted for~\cite{BBFS93,FMS93} the
result is
\begin{equation}
\sigma(b^2)={\pi^2\over 3}\left[ b^2\alpha_s(Q^2)xG_{T}
(x,Q^2)\right]_{x=Q^2/s,Q^2 \simeq 15/b^2} \ ,
\label{eq:9c}
\end{equation}
where $b$ is the transverse distance between the quark $q$ and the
antiquark $\bar q$ and $G_T(x,Q^2)$ is the gluon distribution in the
target $T$ calculated within this approximation. In this equation the $Q^2$
evolution and the small $x$
physics are properly taken into account through the gluon
distribution.
To
derive similar  equation in the  leading
$\alpha_s \ln {Q^2\over \Lambda_{QCD}^2}$ approximation one
should account for all hard processes including diagrams where (anti)quarks in
the box diagram
with   production of
 one hard gluon.
The final result has the same form as
eq.(\ref{eq:9c}), but with $G_N(x,Q^2)$ calculated in
the  leading  $\alpha_s \ln {Q^2\over \Lambda_{QCD}^2}$ approximation. It
also contains a
small contribution due to sea quarks.
Eq.(\ref{eq:9c}) accounts for the contribution of quarks $Q$ whose masses
satisfy the condition:
$l_c = {2 q_0 \over {4 m_Q^2 +Q^2}} \gg r_N^2$.
 The estimate $Q^2 \approx {15 \over b^2}$ was obtained
in \cite{FKS} by numerical analysis
of the $b$-space representation of the
cross section of the longitudinally polarized photon, $\sigma_L$,
 and requiring
that $G_T$ is conventional gluon
distribution calculated in the leading
 $\alpha_s \ln {Q^2\over \Lambda_{QCD}^2}$ approximation.

There is a certain similarity between
equation~(\ref{eq:9c}) and the two gluon exchange model of
F. Low~\cite{Low} and S. Nussinov~\cite{Nussinov}, as well as the
constituent quark 2 gluon exchange model of J. Gunion and D.
Soper~\cite{GS}.  The factor $b^2$ which is present in the QCD
expression~(\ref{eq:9c}) for the cross section is also present in
these models.  The major
qualitative
distinction between the results of QCD
calculations and expectations
  of the two gluon exchange models is
that the nonperturbative QCD physics  is accounted for in
equation~(\ref{eq:9c}) through experimentally measured  quantities -
the gluon and the sea quark distributions
  The latter are  particularly
relevant for the fast increase of the cross section at small $x$, for the
increase of
leading twist nuclear shadowing with decreasing $x$,
for the seemingly slow decrease with $Q^2$ of higher twist processes.
All those effects are characteristic for QCD as a
nonabelian
gauge quantum field
theory which predicts an increase of parton densities in hadrons with
$\frac{1}{x}$ in contrast to quantum mechanical models of hadrons.
In QCD the inelastic cross section for the collision of a sufficiently
energetic small size, colorless two gluon configuration off any target
is \cite{AFS}
\begin{equation}
\sigma(b^2)={3\pi^2\over 4}\left[ b^2\alpha_s(Q^2)xG_{T}
(x,Q^2)\right]_{x=Q^2/s, Q^2=\lambda/b^2} \ ,
\label{eq:9d}
\end{equation}
where the parameter $\lambda$ is likely to be similar to the one
present in the case of scattering of a $q \bar q$ pair off a target.
The difference compared to equation~(\ref{eq:9c}) is in the factor $9/4$
which follows from the fact that gluons belong to the octet representation
of the color group $SU(3)_c$ while quarks are color triplets.

\section{Electroproduction of vector mesons in QCD.}

One of the examples of a new kind of hard processes calculable in QCD
is the coherent electroproduction of vector mesons off a target T,
\begin{equation}
\gamma^* + T \rightarrow V+T \ ,
\label{eq:10c}
\end{equation}
where $V$ denotes any vector meson ($\rho,\omega,\phi, J/\Psi$) or
its excited states.

The idea behind the calculation of hard diffractive processes is that
when $l_c={1\over 2m_N x}$ exceeds the diameter of
the target, the virtual photon transforms into a hadron component well
before reaching the target and the final vector meson $V$ is formed
well past the target. The hadronic configuration of the final state is
a result of a coherent superposition of all those hadronic
fluctuations of the photon of mass $M$
that satisfy equation ${2 l_c \over (1 +M^2/Q^2} \gg r_N$.  Thus,
as in the more familiar leading twist deep inelastic processes, the
calculation should take into account all possible hadronic
intermediate states satisfying this condition.  The use of
completeness over diffractively produced intermediate hadronic states
allows to express the result in terms
distributions of bare
of quarks and gluons as in the
case of other hard processes.  The matrix element of electroproduction
of a vector meson $A$ can be written as a convolution of the light cone
wave function of the photon $\psi^{\gamma^* \rightarrow |n\rangle}$ ,
the scattering amplitude for the hadron state $|n \rangle$,$~A (nT)$,
and the wave function of the vector meson $\psi_{V}$
\begin{equation}
A= \psi^{* ~\gamma^*  \rightarrow |n\rangle}  \otimes
A (nT) \otimes \psi_{V} \ .
\label{eq:11c}
\end{equation}
In the case of a longitudinally polarized photon with high $Q^2$ the
intermediate state $|n\rangle$  is  a $q\bar q$ pair. As was
mentioned in the previous chaptersection, it  can be demonstrated
by direct calculations that the  contribution of higher Fock state
components  and soft physics are suppressed by a factor $1 \over Q^2$
and/or powers of $\alpha_s$. The proof of this result resembles
the calculation of the total cross section for the deep inelastic scattering
in QCD. The situation is qualitatively different in  the case of a
transversely polarized photon due to the singular behavior of the vertex
$\gamma^*_T \rightarrow q\bar q$ when one of the partons carries a small
fraction of the photon momentum. In this case  soft
and hard physics compete in a wide range of $Q^2$.

To understand the applicability of PQCD for the process discussed
above it is convenient to perform the Fourier transform of the
amplitude into the impact parameter space which leads to
\begin{eqnarray}
 A \propto Q\int b^2 x G_T(x,b^2)
K_0\left(Qb\sqrt{z(1-z)}\right)    \nonumber \\
\psi_{V}(z,b) d^2\!b
z(1-z) dz \ ,
\label{brep}
\end{eqnarray}
where $z$ denotes the fraction of the photon momentum carried by one
of the quarks. Here
\begin{equation}
\psi^{\gamma^*_{L}}\propto z(1-z)Q
K_0\left(Qb\sqrt{z(1-z)}\right) \ ,
\end{equation}
where $K_0$  is the Hankel function of an imaginary argument.
To estimate which values of $b$ dominate in the integral we
approximate $\psi_V(z,k_t)$ by ${z(1-z) \over (k_t^2+\mu^2)^2}$ which
corresponds to $\psi_{V}(z,b)\propto  z(1-z)b K_1(\mu b)$.
We vary  $\left< k^2_t\right>^{1/2} = {\mu \over \sqrt{2}}$
 between 300 and 600 MeV/c.

 In  the case of $\sigma_L$ the average transverse size
$\left< b \right> \simeq 0.25 {\rm~fm} $ for $Q^2 = 10 {\rm~GeV}^2, x \sim
10^{-3}$ and decreases at larger $Q^2$ approximately as
$ 0.3 fm{3 GeV \over Q}$ \cite{FKS}. It also weakly decreases with decreasing
   $x$.  The increase of $G_T(x,b^2)$, in equation~(\ref{brep}) with
   decreasing $b$ substantially contributes to the decrease of $\left< b
   \right>$.  In the case of a transversely polarized $\gamma^*$ the
   contribution of large $b$ is not suppressed since
\begin{equation}
  \psi^{\gamma^*_{T}}\propto {\partial \over \partial b_{\mu}}
K_0\left(Qb\sqrt{z(1-z)}\right) \ .
\end{equation}
and therefore the contribution of the kinematical region $z
\rightarrow 0$ and $z \rightarrow 1$ where nonperturbative QCD
dominates is not suppressed.

It is worth noting that $\left< b \right>$ contributing in the
calculation of $\sigma_L$ -- $\left< b (Q^2=10 {\rm~GeV^2})
\right>_{\sigma_L} \simeq 0.25$ fm is similar to that in the
electroproduction of vector mesons $\left< b \left(Q^2=10 {\rm~GeV^2}\right)
\right>_{\gamma^*_L \rightarrow \rho} \simeq 0.35$ fm. However for larger
$Q^2$ the difference between the two values increases and reaches a
factor of 2 for $Q^2 \sim 100$ GeV$^2$.

It can be shown that under certain kinematical conditions the
interaction of a $q\bar q$ pair with the target is given by
equation~(\ref{eq:9c}). In the leading order in $\alpha_s \ln x
\ln{ Q^2\over \Lambda_{QCD}^2}$
the leading Feynman
diagrams for the process under consideration are a hard quark box
diagram with two gluons attached to it and convoluted with the
amplitude for the gluon scattering off a target.

One can consider the same process in the leading $\alpha_s \ln{
Q^2\over \Lambda_{QCD}^2}$ approximation.  In this case one has to
include also the diagrams where one hard quark line is substituted by
the gluon line. This leads to an extra term $ \propto S_T(x,Q^2)$ in
equation~(\ref{eq:9c}) and allows to treat the parton distributions in
equation~(\ref{eq:9c}) with $\alpha_s \ln{ Q^2\over \Lambda_{QCD}^2}$
accuracy which is more precise than the original leading $\alpha_s
\ln x
\ln{ Q^2\over \Lambda_{QCD}^2}$ approximation in equation~(\ref{eq:9c}).

Since Feynman diagrams are Lorentz invariant it is possible to
calculate the box part of the diagram in terms of the light-cone wave
functions of the vector meson and the photon and to calculate the
bottom part of the diagram in terms of the parton wave function of the
proton.  This mixed representation is different from the QCD improved
parton model which only uses the light-cone wave function of the
target.

The next step is to express this amplitude
through the parton distributions in the target.
The calculation of the imaginary part of the relevant
Feynman diagram
shows that the fractions of the target momentum carried by the
exchanged gluons $x_i$ and $x_f$ are not equal,
\begin{equation}
x_i-x_f=x,~~~ for~~~ M_V^2 \ll Q^2
\label{eq:12c}
\end{equation}
We
neglect terms ${\cal O} ({l_t^2\over Q^2})$ as compared to~1,
with $l_t$ the transverse momentum of the exchanged gluons.
Within the QCD leading logarithmic approximation
\begin{equation}
\alpha_s \ln{ Q^2\over \Lambda_{QCD}^2} \sim 1
\label{eq:13c}
\end{equation}
or
\begin{equation}
\alpha_s \ln x \ln{ Q^2\over \Lambda_{QCD}^2} \sim 1
\label{eq:13d}
\end{equation}
 when terms $\sim \alpha_s$ are neglected, the difference between $x_i$
and $x_f$ can be neglected and the amplitude of the $q \bar q$ interaction
with a target is given by
equation~(\ref{eq:9c})~\cite{BBFS93,FMS93,Brod94}.

We are now able to calculate the cross section for the production of
longitudinally polarized vector meson states when the momentum
transferred to the target $t$ tends to zero~\cite {Brod94},
but $Q^2 \rightarrow \infty $ \footnote{In the paper of Brodsky et at
 \cite{Brod94} the factor 4 in eq.(\ref{eq:14c}) has been missed. We
 are indebted to Z.~Chan and
  A.~Mueller
for pointing this out.}

\begin{eqnarray}
& & \left. {d\sigma^L_{\gamma^*N\rightarrow VN}\over dt}\right|_{t=0} =
\nonumber \\
& & {12\pi^2\Gamma_{V \rightarrow e^{+}e^-}
m_{V}\alpha_s^2(Q)\eta^2_V~I_V(Q^2)^2
\over \alpha_{EM}Q^6N_c^2}  \nonumber \\
& & |(xG_T(x,Q^2) + i{\pi\over2}
{d \over d\ln x} xG_T(x,Q)|^2.
\label{eq:14c}
\end{eqnarray}
$\Gamma_{V \rightarrow e^{+}e^-}$ is the decay width of the vector
meson into $e^+e^-$.
The parameter $\eta_V$ is defined as
\begin{equation}
\eta_V\equiv {1\over 2}{\int{dz\over z(1-z)} \Phi_V(z)\over
\int dz \Phi_V(z)}  \ ,
\label{eq:15c}
\end{equation}
where $\Phi_V$ is the light cone wave function of the vector meson.
At large $Q^2$ equation~(\ref{eq:14c}) predicts a $Q^2$ dependence of
the cross section which is substantially slower than $1/Q^6$ because
the gluon densities at small $x$ fastly increase with $Q^2$.
Numerically, the factor $\alpha_s^2(Q^2) G^2(x,Q^2)$
in equation~(\ref{eq:14c}) is
$\propto Q^n$ with $n \sim 1  $.
An additional $Q^2$ dependence of the cross section arises from the
transverse momentum overlapping intergral between the light-cone wave
function of the $\gamma^*_L$ and that of the vector meson~\cite{FKS},
expressed through the ratio $I_V(Q^2)$
\begin{equation}
I_V(Q^2)= {\int_0^1 {dz \over z(1-z)}\int_0^{Q^2}d^2k_t
{Q^4 \over \left[ Q^2 +{k_t^2+m^2 \over z(1-z)}\right]^2} \psi_V(z,k_t)
\over  \int_0^1 {dz \over z(1-z)}\int_0^{Q^2}d^2k_t \psi_V(z,k_t)}.
\end{equation}
In ref.~\cite{Brod94} it was assumed that $I_V(Q^2)=1$ as for $Q^2
 \rightarrow \infty $ the ratio $I_V(Q^2)$ tends to 1.  But for
 moderate $Q^2$ this factor is significantly smaller than 1.  For
 illustration we estimated $I_V(Q^2)$ for the following vector meson
 wave function: $\psi_V^{(1)}(z,k_t^2)={ c z(1-z) \over
 (k_t^2+\mu^2)^2}$.  The momentum dependence of this wave function
 corresponds to a soft dependence on the impact parameter $b$ - $exp(-\mu
 b)$ in coordinate space.  We choose the parameter $\mu$ so that
 $\left<k^2_t\right> ^{1/2} \in 0.3 \div 0.6~GeV/c$.

Our numerical studies show that the inclusion of the quark transverse
momenta leads to several effects:
\begin{itemize}
\item Different $k_{T}$ dependence of $\psi_V$ leads to somewhat different
$Q^2$ dependence of
$I_V(Q^2)$. Thus
measuring of $Q^2$ dependence of
electroproduction of vector mesons may become an
effective way of probing $k_t$-dependence of the light-cone $q \bar q$
wave function of vector mesons.

\item The $Q^2$ dependence of $I_V$ for production of vector
mesons build of light quarks $u,d,s$ should be very similar.

\item For  electroproduction  of charmonium states
where $\mu_c \sim \mu {m_{J\Psi }\over m_{\rho}}$ the asymptotic
formula should be only valid for extremely large $Q^2$.
\end{itemize}
The NMC data~\cite{NMC1} and the  HERA data~\cite {ZEUSb} on
diffractive electroproduction of $\rho$ mesons are consistent with
several predictions of equation~(\ref{eq:14c}):
\begin{itemize}
\item a fast increase
with energy of the cross section for electroproduction of vector
mesons  (proportional to $x^{-0.8}$ for $Q^2
=10$~GeV$^2$) (figure 1 \cite{FKS})~\footnote{This fast increase with
decreasing $x$ is
  absent in the non--perturbative two--gluon exchange model of
  Donnachie and Landshoff~\cite{DL1} which leads to a cross section
  rising as $\sim x^{-0.14}$ at $t=0$ and to a much weaker increase of
  the cross section integrated over t.};
\item the dominance of the
longitudinal polarization ${\sigma_L \over \sigma_T}\propto Q^2$;
\item
 the absolute magnitude of the cross section within the uncertainties
 of the gluon densities and of the $k_t$ dependence of the wave
 functions (figure 1)
\item the $Q^2$ dependence of the cross section for
$Q^2 \sim 10 {\rm~GeV^2}$ which can be parameterized as $Q^{-n}$ with $n
\sim 4$.  The difference of $n$ from the asymptotic value of 6 is due
to the $Q^2$ dependence of $\alpha_s^2(Q^2)G_N^2(x,Q^2)$ and of
$I_V^2$ which are equally important in this $Q^2$ range.
\end{itemize}
\begin{figure}
\epsfxsize=15cm
\centerline{\epsffile{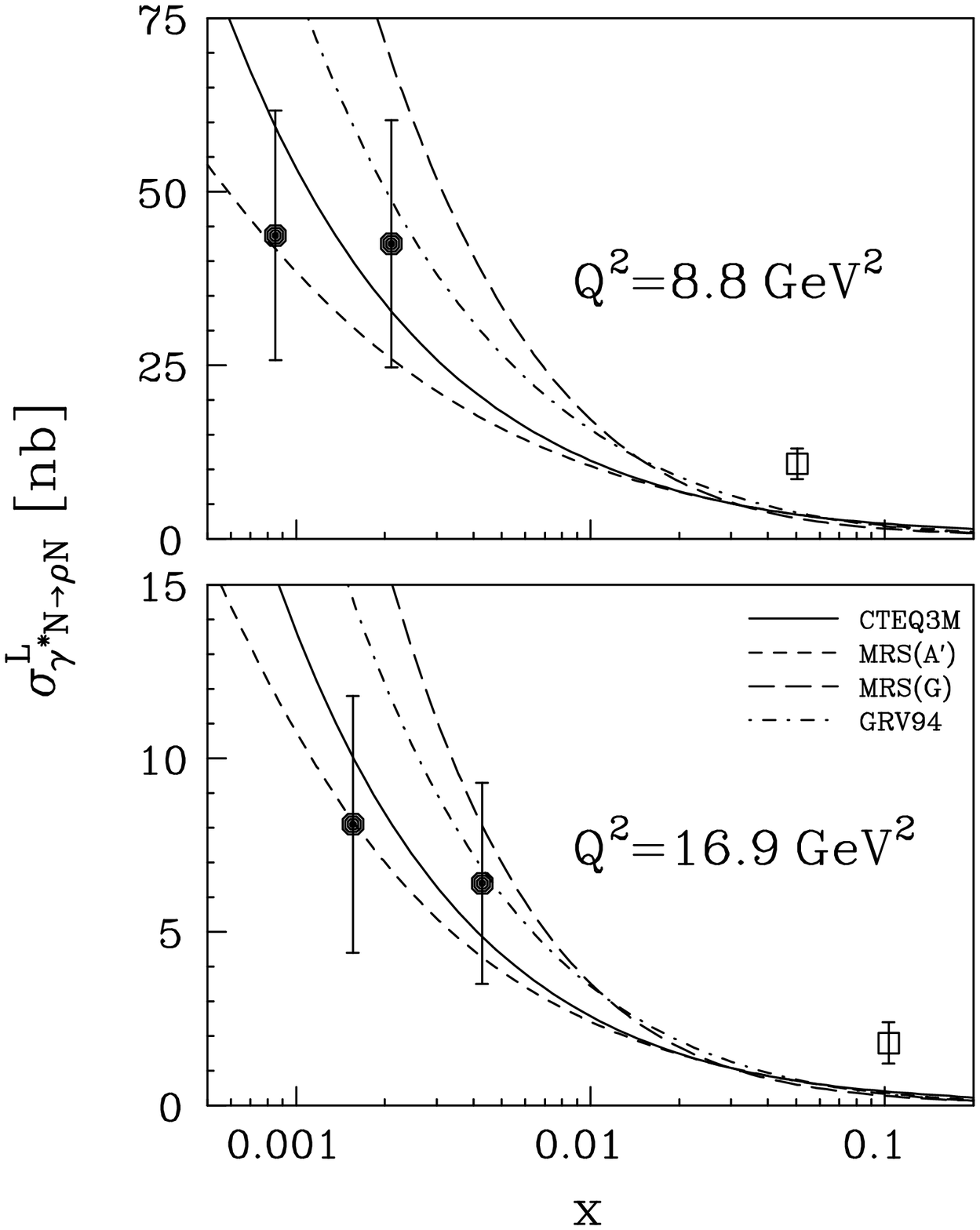}}
\vspace{-1.5cm}
\protect{\caption{The total longitudinal cross section, $\sigma_{\gamma^* N
\rightarrow
\rho N}^L$, calculated from Eq. (\protect\ref{eq:14c}) for several recent
parameterizations of the gluon density in comparison with experimental data
from ZEUS \protect\cite{ZEUSb} (full circles) and NMC \protect\cite{NMC1}
(squares). Typical parameters
for the $\rho$-meson wave functions as discussed above are taken
($\left<k_t^2\right>^{1/2} =0.45 GeV/c$).  We set
$\eta_V=3$ and parameterize the dependence of the differential cross section
on the momentum transfer in exponential form with $B \approx 5$ GeV$^{-2}$.
Note that a change of $T^2(Q^2)$ in the range
corresponding to $\left<k_t^2\right>^{1/2}$ between
0.3 GeV/c and 0.6 GeV/c introduces
an extra scale uncertainty of $0.7 \div 1.4$.
}}
\end{figure}

We discussed above (see also section 8) that the perturbative regime
should dominate in the production of transversely polarized vector
mesons as well, though at higher $Q^2$. This may
be
manifested   in
the $x$-dependence of the ratio ${\sigma_L \over \sigma_T}$ for fixed
$Q^2$.  At intermediate $Q^2 \sim 10 {\rm~GeV^2}$ where hard physics
already dominates in $\sigma_L$, $\sigma_T$ may still be dominated by
soft nonperturbative contributions.  For these $Q^2$ the ratio should
increase with decreasing $x$ $\sim x^2G_N^2(x,Q^2)$.  At sufficiently
large $Q^2$ where hard physics dominates for both $\sigma_L$ and
$\sigma_T$ the ratio would not depend on $x$.

The t dependence of the cross section is given by the square of the
two gluon form factor of the nucleon $G_{2g}(t)$.  Practically no t
dependence should be present in the block of $\gamma^*$ gluon
interaction for $-t\ll Q^2$.  Thus {\bf the t dependence should be
  universal for all hard diffractive processes}.  Experimentally the
data on diffractive production of $\rho$ mesons for $Q^2 \ge
5{\rm~GeV}^2$~\cite{NMC1}, on photoproduction of $J/\Psi$
mesons~\cite{psi1} and even on neutrinoproduction of $D^*_s$
mesons~\cite{Asratian} show a universal $t$ behavior corresponding to
$G_{2g}^2(t) =\exp (Bt)$ with$ B \approx 4 \div 5{\rm~GeV}^{-2} $.
A certain weak increase of $B$ is expected with increasing incident
energy due to the so called Gribov diffusion~\cite{Gribovdif}, but
this effect is expected to be much smaller than for soft processes.
However in the limit $Q^2=const$ and $s \rightarrow \infty$ it is
natural to expect an onset of a soft regime, which is characterized
both by a slowing down of the increase of the cross section with
increasing $s$ and by a faster increase of the slope $B$ with $s$,
\begin{equation}
 {\partial \ln B \over \partial \ln s}_ {| s \rightarrow \infty,
Q^2=const} \approx \alpha^{\prime}_{soft} \approx 0.25 GeV^{-2}.
\label{slope1}
\end{equation}
For further discussion see section 10.

We want to point out that for $M_X^2\ll Q^2 $, the effect of
QCD radiation  is small.
This is because bremsstrahlung corrections
due to radiation of hard quarks and gluons
are controlled by the parameter $\alpha_s \ln{x_i\over x_f}$ which is
small since in the reaction considered here $x_i \sim x_f$.  This
argument can be put on a formal ground
within the double logarithmic approximation when only terms $\sim
\alpha_s \ln {1 \over x} \ln {Q^2\over \lambda^2}$ are taken into
account.
One can consider a
more traditional approximation where terms $\simeq \alpha_s
\ln{Q^2\over\Lambda_{QCD}^2}$ are taken into account but terms $\simeq
\alpha_s$ are neglected.
Within these approximations it is legitimate to neglect the
contribution
of
the longitudinal momentum as compared to the
transverse one. This is a special property of small $x$ physics. Thus
the difference between $x_i$ and $x_f$ leads to an insignificant
correction.

Formula~(\ref{eq:14c}) correctly accounts for nonperturbative
physics and for the diffusion to large transverse distances characteristic
for Feynman diagrams, because  in contrast to the naive applications of the
BFKL Pomeron the diffusion of small size
configurations to large transverse size is not neglected.

  Electroproduction of $J/\Psi$ mesons has been
calculated in the whole $Q^2$ range
in~\cite{Ryskin}
within the leading $\alpha_s \ln x$ approximation
of QCD for the interaction with a target
and the nonrelativistic charmonium
quark model for $J/\Psi$-meson wave function.
If we would apply  (eq.~(\ref{eq:14c})) at $Q^2=0$
the result of ref.~\cite{Ryskin}
coincides with the nonrelativistic limit of our result if $I_V$ is assumed
to be equal 1.
At the same time the inclusion of the transverse momentum distribution of
$c$ quarks in the $J/\Psi$ wave function significantly suppresses the
cross section of the diffractive electroproduction of $J/\Psi$ mesons
for $Q^2 \le m^2_{J/\Psi}$.  In particular,
account of the quark Fermi motion within the model of ref.\cite{Ryskin}
using realistic charmonium models
lead to suppression  of photoproduction cross section
by a factor $4 \div
 8$ depending on the
charmonium
model (see discussion in ref.\cite{FKS}).
Remember that transverse distances essential in the photoproduction
of the $J/\Psi$ meson are $\sim {3\over m_c}$ which
are
comparable to the
 average interquark distance in the $J/\Psi$ wave function.
 Since the energy dependence of diffractive photoproduction of
 $J/\Psi$ is consistent with pQCD prediction of \cite{Ryskin}
 the disagreement with the absolute prediction may indicate an important
 role for the interaction with interquark potential.
Note that in the limit
where it is possible to justify the application of PQCD (eq.~(\ref{eq:14c}))
($m^2_{J/\Psi}\ll Q^2 $) it is necessary to use distribution
of bare c quarks within ${J/\Psi}$ meson instead of charmonium wave
functions  to account for the
screening of color fields of c quarks.

Another interesting process which can be calculated using the
technique discussed above is the production of vector mesons in the
process $\gamma_L^* +p \rightarrow V +X$ in the triple Reggeon limit
when $-t \ge$~few~GeV$^2$ and $ -t \ll Q^2$.  In this kinematical
domain the dominant contribution is due to the scattering of the two
gluons off a parton of the target $g+g+{\rm parton} \rightarrow {\rm
  parton}$.  To avoid the uncertainties related to the vector meson
wave function it is convenient to normalize the cross section of this
process to that of the exclusive vector meson production at $t=0$
\cite{F94,AFS}
\begin{eqnarray}
 {{d\sigma^{\gamma_L^* +p \rightarrow V +X}\over dt}\over
\left.{d \sigma^{\gamma_L^* +p \rightarrow V +p}\over dt}
\right|_{t=0}}
=   {9\over 8 \pi} \alpha_S^2 \left| \ln{Q^2\over k^2} \right|^2\times
\ \ \ \ \ \ \nonumber \\
%&   &
{\int\limits^1_y
\left[ G_p(y',k^2)
+{32 \over 81}S_p(y',k^2) \right] dy' \over \left[ xG_p(x,Q^2) \right] ^2}\ ,
\label{larget}
\end{eqnarray}
where $S_p$ is the density of charged partons in the proton, $\nu=2m_N
q_o$, $x=Q^2/\nu$, $k^2=-t$, $y=-t/{2(q_o-p_{Vo})}m_N$ with $p_{Vo}$
the energy of the vector meson and all variables are defined in the
nucleon rest frame.

It follows from equation~(\ref{larget}) that the cross section of the
process $\gamma_L^* +p \rightarrow V +X$ should decrease very weakly
with $t$ and therefore it is expected to be relatively large at
$-t\sim {\rm few~ GeV}^2$.  Similarly to the approach taken
in~\cite{FS89,MuTan} one can easily improve equation~\ref{larget} to
account for leading $\alpha_s \ln x$ terms.

Equation~(\ref{larget}) is a particular case of the suggestion (and of
the formulae) presented in reference~\cite{FS89}, that semi--exclusive
large t diffractive dissociation of a projectile accompanied by target
fragmentation can be expressed through the parton distributions of the
target. The advantage of the process considered here as compared to
the general case is the possibility to prove the dominance of hard
PQCD physics for a longitudinally polarized photon as
the
projectile and
the lack of t dependence in the vertex $\gamma^* +g\rightarrow g + V$.
These advantages allow to calculate the cross section without free
parameters.

 Production of transversely polarized vector mesons by real or virtual
photons in the double diffractive process $\gamma_T +p \rightarrow V
+X$ has been calculated recently within the approximation of the BFKL
Pomeron in~\cite{FR}.  The calculation was performed in the triple
Reggeon limit for large t but $s\gg -t$.  Contrary to reactions
initiated by longitudinally polarized photons this calculation is
model dependent; the end point nonperturbative contribution to the
vertex $\gamma_T^* +g \rightarrow g+V$, and therefore to the whole
amplitude, leads to a contribution which is not under theoretical
control.  This problem is familiar to the theoretical discussions of
high $Q^2$ behavior of electromagnetic form factors of hadrons.

 Perturbative QCD predicts also approximate restoration of $SU(3)$
symmetry in the production of vector mesons at large $Q^2$ and significant
enhancement of the production of $J/\Psi$ meson as compared to
$SU(4)$:
\begin{equation}
\rho^o : \omega : \phi
: J/\Psi =9 : 1 : (2*1.0) : (8*1.5).
\label{eq:19c}
\end{equation}
This prediction is valid for $Q^2 \gg m_V^2$ only. Pre--asymptotic
effects are important in the large $Q^2$ range. They significantly
suppress the cross section for production of charmonium states (see
above discussion). Thus the value of the $J/\Psi /\rho$ ratio would be
significantly below the value given by eq.(\ref{eq:19c}) up to very
large $Q^2$. For example the suppression factor is $\sim 1/2$ for $Q^2
\sim 100 GeV^2$ \cite{FKS}.
At the same time it is likely to change very little the predictions for $\rho,
\omega, \phi $-meson production,  since the masses of these
hadrons are quite close and their $q\bar q$ components should be very
similar.

At very large $Q^2$ the $q\bar q
$ wave functions of all mesons converge to a universal asymptotic wave
function with $\eta_V=3$. In this limit further enhancement of the
heavy resonance production is expected
\begin{equation}
\rho^o : \omega : \phi : J/\Psi =9 : 1  : (2*1.2) : (8*3.4) \ .
\label{eq:20c}
\end{equation}
It is important to investigate these ratios separately for the
production of longitudinally polarized vector mesons where hard
physics dominates and for transversely polarized vector mesons where
the interplay of soft and hard physics is more important.

Equation~(\ref{eq:14c}) is applicable also for the production of
excited vector meson states with masses $m_V$ satisfying the condition
that $ m_V^2\ll Q^2 $.  In this limit it predicts comparable
production of excited and ground states.  There are no estimates of
$\eta_V$ for these states but it is generally believed that for
$\rho'$, $\omega'$ and $\phi'$ it is close to the asymptotic value,
and as a rough estimate, we will assume that $\eta_V=\eta_{V'}$. Using
the information on the decay widths from the Review of Particle
Properties~\cite{RPP} we find that
\begin{eqnarray}
\rho(1450):\rho^o \approx \omega(1420):\omega  \approx 0.3 \nonumber \\
   \rho(1700):\rho^o \approx \omega(1600):\omega  \approx  1.0 \nonumber \\
\phi(1680) :   \phi \approx 0.6, ~
 \Psi' :J/\Psi  \approx  0.5 \ .
\label{eq:21c}
\end{eqnarray}
In view of substantial uncertainties in the experimental widths of
most of the excited states and substantial uncertainties in the values
of $\eta_{V'}$ and the ratio ${I_{V'} \over I_V}$
these numbers can be considered as good to about a
factor of 2.
The case of $ \Psi'$ where $\Gamma_V $ is well known is less
ambiguous. In this case estimates using charmonium models indicate a
significant suppression as compared to the asymptotic estimate up to
$Q^2 \sim 20 GeV^2$ where this suppression is $\sim 0.5$ \cite{FKS}.
In spite of these uncertainties it is clear that
 a substantial production of excited resonance states is
expected at large $Q^2$ at HERA. A measurement of these reactions may
help to understand better the dynamics of the diffractive production
as well as the light-cone minimal Fock state wave functions of the
excited states. It would allow also to look for the second missing
excited $\phi $ state which is likely to have a mass of about 1900 MeV
to follow the pattern of the $\rho,~\omega,~J/\Psi$ families.

The predicted
relative
yield of the excited states induced by virtual photons is
expected to be higher than for real photons.
Another interesting QCD effect is that the ratio of the cross section
for the diffractive production of excited and ground states of vector
mesons should increase with decreasing $x$ and
increase
$Q^2$. This is because
the energy denominator - ${1 \over \left({m_q^2+k^2_t \over z(1-z)} -
m_{V'}^2\right)}$, relevant for the transition $V \rightarrow q \bar
q$ (with no additional partons) should be large and positive. Thus
the heavier the excited state, the larger Fermi momenta should be
important. Thus the gluon distributions should enter at larger
virtualities in the case of $V'$ production.

To summarize, the investigation of exclusive diffractive processes
appears as the most effective method to measure the minimal Fock
$q\bar q$ component of the wave functions of vector mesons and the
light-cone wave functions of any small mass hadron system having
angular momentum 1. This would be very helpful in expanding methods of
lattice QCD into the domain of high energy processes.

\section{Electroproduction of photons.}

The diffractive process $\gamma^* + p \rightarrow \gamma + p$ offers
another interesting possibility to investigate the interplay between
soft and hard physics and to measure the gluon distribution in the
proton.  We shall consider the forward scattering in which case only
the transverse polarization of the projectile photon contributes to
the cross section.  This follows from helicity conservation.  In this
process, in contrast to reactions initiated by longitudinally
polarized highly virtual photons, soft (nonperturbative) QCD physics
is not suppressed. As a result, theoretical predictions are more
limited. Within QCD one can calculate unambiguously only the
derivative of the amplitude over $\ln{Q^2\over Q_o^2}$ but not the
amplitude itself. However for sufficiently small $x$ and large $Q^2$,
when $\alpha_s(Q_o^2) \ln{Q^2\over Q_o^2} \ln x$ is large, PQCD
predicts the asymptotic behavior of the whole amplitude.

It is convenient to decompose the forward scattering amplitude for the
process $\gamma^* + p \rightarrow \gamma + p$ into invariant structure
functions in a way similar to the case of deep inelastic
electron-nucleon scattering. Introducing the invariant structure
function $H(x,Q^2)$, an analogue of $F_1(x,Q^2)$ familiar from deep
inelastic electron scattering off a proton, we have \cite{AFS}
\begin{equation}
\left. {d\sigma  \over dt}^{\gamma^*+N \rightarrow \gamma + N}
\right| _{t=0} =
\pi \alpha_{em}^2 {H(x,Q^2)^2\over {s^2}} \ .
\label{eq:26c}
\end{equation}

When $Q^2$ is sufficiently large, QCD allows to calculate the $Q^2$
evolution of the amplitude in terms of the parton distributions in the
target. As in the case of deep inelastic processes it is convenient to
decompose $H(x,Q^2)$ in terms of photon scattering off flavors of type
$i$
\begin{equation}
H(x,Q^2)=\sum\limits_{i}  e_i^2 h_i(x,Q^2) \ ,
\label{eq:27c}
\end{equation}
where the sum runs over the different flavors $i$ with electric charge
$e_i$. It is easy to deduce the differential equation for $h_i$, the
analogue of the evolution equation for the parton distributions.
\begin{eqnarray}
  {dh_i(x,Q^2) \over d\ln{Q^2 }} = {\alpha_{s}(Q^2)\over 2\pi} \int
  {dz\over z} \left[ P_{qg}\left( {x\over z}\right) G_p(z,Q^2) +
  \nonumber \right. \\ \left.  P_{q q}\left( {x\over z}\right) q_i(z,
  Q^2)\right] \left[ 1 + {x\over z}\left( 1-{x\over z}\right) \right]
  +{\cal O}(\alpha_s^2) \ .
\label{eq:28c}
\end{eqnarray}
Here $P_{qq}$ and $P_{qg}$ are the splitting functions of the GLDAP
evolution equation~\cite{GLDAP} . The factor $ 1 + {x\over z}\left(
1-{x\over z} \right)$ takes into account the difference of the
virtualities of the initial and final photon. The solution of this
equation is
\begin{eqnarray}
 h_i(x,Q^2)  =  h_i(x,Q_{0}^2) +{\alpha_{s}(Q^2)\over 2\pi}
\int\limits_{\ln Q_o^2}^
{\ln Q^2}d\ln Q_1^2 \int\limits_{x}^{1} {dz\over z} \nonumber \\
\left[ P_{qg}\left( {x\over z}\right) G_p(z,Q_1^2) +
P_{q%\bar
  q}\left( {x\over z} \right) q_i(z, Q_1^2) \right]\nonumber \\
\left[ 1 + {x\over
  z}\left( 1-{x\over z}\right) \right] +{\cal O} (\alpha_s^2) \ .
\label{eq:29c}
\end{eqnarray}
Usually it is assumed that the soft components of the parton
distributions increase at small $x$ more slowly than the hard ones. If
this is the case, at sufficiently small $x$, in the leading $\alpha_s
\ln x$ approximation, the first term in equation~(\ref{eq:29c}) can be
neglected.  As a result one can obtain the asymptotic formula for the
whole $H(x,Q^2)$ and not only for its derivative.

Similarly to the case of electroproduction of photons it is not
difficult to generalize the $Q^2$ evolution equation to the amplitude
for the diffractive production of transversely polarized vector mesons.
One of the consequences of this evolution equation is that, at
asymptotically large $Q^2$ and small $x$, the production cross section
has the same dependence on the atomic number of a target as in the
case of longitudinally polarized vector mesons.

\section{Coherent Pomeron.}

It is interesting to consider high-energy hard processes in the
diffractive regime with the requirement that there is a large rapidity
gap between the diffractive system containing the high $p_t$ jets and
the target which can remain either in the ground state or convert to a
system of hadrons. In PQCD such a process can be described as an
exchange of a hard gluon accompanied by a system of extra gluons which
together form a color neutral state.  It was predicted~\cite{FS89}
that such processes should occur in leading twist. ( Note that in
reference~\cite{MR} it was stated that this process should rather be a
higher twist effect.  This statement was due to some specific assumptions
about the properties of the triple Pomeron vertex).

The simplest example is in the triple Reggeon limit the production of
high $p_t$ jets in a process like
\begin{equation}
h + p \rightarrow jet_1 + jet_2 +X +p
\label{eq:30c}
\end{equation}
where the final state proton carries practically the whole momentum of
the initial proton.  The initial particle can be any particle
including a virtual photon.  To probe the new PQCD hard physics the
idea~\cite{FS89} is to select a final proton with a large transverse
momentum $k_t$. One can demonstrate that this selection tends to
compress initial and final protons in small configurations at the
moment of collision. In this case the use of the PQCD two gluon
exchange or two--gluon ladder diagrams becomes legitimate. A
nontrivial property of these processes is a strong asymmetry between
the fractions of the target momentum carried by the two gluons (the
contribution of the symmetric configurations is a higher twist effect
with the scale determined by the invariant mass of the produced two
jets~\cite{CFS93}). Thus one expects gluon bremsstrahlung to play a
certain role \cite{F92}.  However since the proton is in a
configuration of a size $\sim {1 \over k_t}$ this radiation is
suppressed by the small coupling constant: $\sim \alpha_s(k_t^2) \ln
({p_t^2\over k_t^2})$.  When $k_t$ tends to 0 this radiation may
suppress significantly the probability of occurrence of events with
large rapidity gaps.

The prediction is that such a process appears as a leading twist
effect \cite{FS89}
\begin{equation}
{d\sigma\over dp_t^2} \sim {1\over p_t^4} \ .
\label{eq:31c}
\end{equation}
This prediction is in an apparent contradiction with a naive
application of the factorization theorem in QCD which states that the
sum of the diagrams with such soft gluon exchanges cancels in the
inclusive cross section. However in reaction~(\ref{eq:30c}) we
selected a certain final state with a white nucleon hence the usual
proof of the factorization theorem does not hold anymore --- there is
no cancelation between absorption and radiation of soft
gluons~\cite{CFS93}.  This conclusion was checked in a simple QED
model with scalar quarks~\cite{SB}.

It was suggested by Ingelman and Schlein~\cite{IS} to consider
scattering off the Pomeron as if the Pomeron were an ordinary particle
and to define parton distributions in the effective Pomeron. In this
language the mechanism of hard interaction in diffraction discussed
above would contribute to the parton distribution in the Pomeron a
term proportional to
  \begin{equation}
 \delta(1-x) \ \ \  {\rm or}  \ \ \ \    {1\over (1-x)} \ \ .
\label{eq:33c}
\end{equation}
This term corresponds to an interaction in which the Pomeron acts as a
whole. Hence the term coherent Pomeron.  In this kinematical
configuration the two jets carry practically all the longitudinal
momentum of the Pomeron.  The extra gluon bremsstrahlung discussed
previously renders the $x$ dependence somewhat less singular at $x
\rightarrow 1$ but the peak should be concentrated at large
$x$~\cite{F92,CFS93,SB}
%%we need here the  reference to last paper of soper and arjun.
There are no other known mechanisms generating
a peak at large $x$.  The recent UA(8) data~\cite{Br} on the reaction
$p +\bar p \rightarrow jet_1 + jet_2 +X +p$, with the proton
transverse momentum in the range $2{\rm~GeV}^2\geq k_t^2\geq
1{\rm~GeV}^2$, seem to indicate that a significant fraction of the two
jet events corresponds to the $x \sim 1$ kinematics.  It is thus
possible that the coherent Pomeron contributes significantly to the
observed cross section~\footnote{ The coherent production of high
  $p_t$ jets by a real photon has been first discussed by Donnachie
  and Landshoff ~\cite{DL2} and then rediscussed in
  reference~\cite{XY}.  This process, discussed in the next section,
  gives a negligible contribution in the kinematic regime
  characteristic for the coherent Pomeron.}.

 The prediction is that the contribution of the coherent
Pomeron to diffractive  electroproduction of dijets
at $p_t^2 \gg Q^2$ should be suppressed by an additional power of $Q^2$
$${d \sigma^{\gamma^* +p \rightarrow 2 jets +X + p} \over dp^2_t} \sim
{1 \over p^4_t} {1 \over Q^2}$$ as compared to
$${d \sigma^{\gamma^* +p \rightarrow 2 jets +X } \over dp^2_t} \sim {1
  \over p^4_t} $$ for other hard processes originating from the hard
structure of the virtual photon.

The complicated nature of the
effective Pomeron should manifested itself in several ways in hard
diffraction~\cite{FS89,CFS93}.\\
(i) There should be a significant suppression of the coherent Pomeron
mechanism at small $t$ due to screening (absorptive) effects since at
small $t$ the nucleon interacts in an average configuration.  This
suppression should be larger for $pp$ scattering than for $\gamma p$
scattering since absorptive corrections increase with the increase of
the total cross section (for $\gamma p$ interaction the VDM effective
total cross section at HERA energies is $\leq 30$ mb).\\
(ii) Due to the contribution of soft physics, the effective Pomeron
structure function as determined from the low $t$ diffractive
processes should be softer than for large $t$ diffraction.

Therefore it would be very important to compare hard diffractive
processes induced by different projectiles and to look for deviations
from the predictions based on the simplest assumption that the Pomeron
has an universal parton distribution~\cite{CTEQ}.

\section{Forward electroproduction of jets.}

Forward diffractive photo and electroproduction of high $p_t$ jets off
a nucleon target (in the photon fragmentation region) $\gamma^* + N
\rightarrow jet_1 +jet_2 +N$ is another promising process to
investigate the interplay of soft and hard physics. We shall confine
our discussion to the kinematical region
\begin{equation}
{-\left<r_N^2\right> t_{min}\over 3}=
\left({Q^2+M_{q\bar q}^2 \over 2q_0}
\right)^2
{\left<r_N^2\right> \over 3} \ll 1,
\label{eq:33c1}
\end{equation}
where
\begin{equation}
 M_{q\bar q}^2 = {(m_q^2 + p_t^2)\over z(1-z)}
\label{eq:34c1}
\end{equation}
is the square of the invariant mass of the  produced $q\bar q$ system, $m_q$
is the mass of quarks and $z$ is the fraction of photon momentum carried
by the $q$ or $\bar q$. In this regime the coherence of the produced hadron
states allows to express the amplitude through the gluon distribution in
the target.

An interesting effect occurs in the photoproduction
 The contribution of a single Feynman diagram with the 2 gluon
exchange in the $t$ channel contains terms $R_1\approx {p_{t~\mu}\over
p^2_t+M^2}$ and $R_2 \approx {m \over p^2_t}$. Here
 is the mass of a bare quark, $M$ can be calculated through $m$
in pQCD but in general accounts for the nonperturbative physics. We omit
constants and $\sigma$ matrixes in this dimensional estimate and
restrict ourselves to the contribution of large $p_t$ only.  A
cancelation occurs when the sum of diagrams is considered. It
accounts for the fact that the sum of diagrams describes the
scattering of a colorless dipole.

   Naively we should expect that after cancelation $R_1$ term
should become $R_1\approx {p_{t~\mu}\over (p_t^2+M^2)^2}$. But in reality
it becomes $R_1\approx {M^2 p_{t~\mu}\over (p_t^2+M^2)^3}
\approx {1 \over p_t^5}$.
$R_2$ term after cancelation in the sum of diagrams becomes
$R_2\approx  {m\over {p_t^4}}$.
            Thus cross section of forward photoproduction
of $q \bar q$ pair $d\sigma/dt dp_t^2$ contains terms:
   $m^2\over p_t^8$ \cite{XY},  ${M^4\over p_t^{10}}$
       and ${M^2 m\over p_t^9}$.

  Since mass of light quark is small it is reasonable to put it 0.
It is not legitimate to put $M=0$. So expected asymptotical behavior is
${M^4\over p_t^{10}}$.
 Thus photoproduction of charm
should dominate hard diffractive photoproduction processes for $ p_t
\geq m_c$ \cite{XY}.

Photoproduction of high $p_t$ jets originating from the
fragmentation of light flavors is predominantly due to next to leading
order processes in $\alpha_s$.

The diffractive electroproduction of dijets seems to be the dominant
process in the region of $M_{q\bar q}^2\leq Q^2$, while in the region
$M_{q\bar q}^2\gg Q^2$ exclusive dijet production is one of many
competing processes contributing to the diffractive sector like
radiation of gluons from quark and gluon lines.

In the approximation when only leading
$\alpha_s \ln{{Q^2\over \Lambda_{QCD}^2}}$ or leading
$\alpha_s \ln x \ln{Q^2\over \Lambda_{QCD}^2}$
 terms are kept,
the  off mass shell effects in the amplitude for the $q\bar q$ interaction
with a target are unimportant. Therefore  the  total cross section  of
diffractive electroproduction of jets by longitudinally polarized
photons can be calculated by applying  the optical theorem for the elastic
$q\bar q$ scattering off a nucleon target and equation~(\ref{eq:9c}) for
the total cross section of $q\bar q$ scattering off a nucleon:
\begin{eqnarray}
\sigma(\gamma_L^* + N \rightarrow jet_1 +jet_2 +N)= \nonumber \\
 {1\over 16\pi B}
\int \psi^2_{\gamma_L^*}(z,b)\cdot(\sigma(b^2))^2 dz d^2b
\label{eq:34d1}
\end{eqnarray}
Here B is the  slope of the two gluon form factor  discussed in section~4
and $\psi_{\gamma_L^*}(z,b)$ is the wave function of the longitudinally
polarized photon.
Essentially the
same equation is valid for the production by
transversely polarized virtual photons of two jets which share equally the
momentum of the projectile photon.

In Ref.\cite{Ryskin2} it has been assumed that diffractive production of
jets off a proton is
dominated by hard physics and that soft physics ia unimportant.
The formulae
obtained under this assumption resembles equation~(\ref{eq:34d1})
but with the
gluon distribution in a target
calculated within the leading  $\alpha_s \ln {1 \over x}$ approximation.
In view of the nontrivial interplay of soft and hard physics of large
longitudinal distances this approach is
difficult to justify
in QCD.
To visualize this point let us consider
the effect
of nuclear shadowing in diffractive electroproduction of jets.  If the
assumption that hard PQCD dominates at each stage of the interaction
were correct, nuclear shadowing should be numerically small and
suppressed by a power of $Q^2$. On the contrary, in QCD at sufficiently small
$x$ and fixed $Q^2$ nuclear shadowing is expected to be substantial
and universal for all hard processes. This conclusion is supported by
current data on nuclear shadowing in deep inelastic processes.

Dijet production has been also considered in the constituent quark
model of the proton~\cite{NZ,NZZ}. In this approach the cross section
for diffraction is expressed through a convolution of the quark
distribution in the virtual photon, the distribution of constituent
quarks in the proton and their interaction cross section. A later
generalization of this model~\cite{NZZ} includes the gluon field of
constituent quarks. In QCD though, hard processes have to be expressed
in terms of bare partons and not constituent ones. This is due to the
use of completeness of the intermediate hadronic states in hard
processes.

Equation~(\ref{eq:34d1}) implies that in this higher twist effect the
contribution of large $b$, that is of the nonperturbative QCD, is
enhanced as compared to the large $b$ contribution to the total cross section.
This result has been anticipated in the pre-QCD times~\cite{BjKogut}
and has been confirmed in QCD~\cite{FS88}. A similar conclusion has
been reached in the constituent quark model~\cite {NZZ} approach
which however ignores characteristic for QCD  increase of parton
distributions at small $b$.
In QCD the hard contribution
may
become dominant only
at rather small $x$ and large $Q^2$.
A similar conclusion has been reached for the
cross section of diffractive processes, calculated in the
approximation of the BFKL Pomeron~\cite{Bartels}, in the triple
Reggeon region when the mass of the produced hadronic system is
sufficiently large $M^2\gg Q^2$.

Note that PQCD diagrams which were found to dominate in the large mass
diffraction~\cite{Bartels} are different from those expected from the
naive application of the BFKL Pomeron~\cite{Ryskin2,NZZ} and lead to
different formulae.

To calculate this process within the more conventional leading
$\alpha_s \ln Q^2$ approximation it is necessary to realize that in
the kinematical region where $M_{q\bar q}^2\sim Q^2$ the fractions of
nucleon momentum carried by the exchanged gluons are strongly
different, $x_{\rm hard} \simeq 2x$ but $x_{\rm soft}\ll x$. This is
qualitatively different from the case of the vector meson production
considered in section 3 in which the two values of $x$ of the gluons
were comparable. This is because in the case of dijet production the
masses of the intermediate states are approximately equal to the mass
of the final state.  As a result of the asymmetry of the two $x$
values the overlap integral between the parton wave functions of the
initial and final protons cannot be expressed directly through the
gluon distribution in the target.  However at sufficiently small $x$
and large $Q^2$, when the parameter ${\alpha_s\over \pi}\ln x
\ln{Q^2\over \Lambda^2} \sim 1$, electroproduction of high $p_t$
dijets can be expressed through the gluon distribution in a target but
in a more complex way. In this particular case the factorization
theorem can be applied after the first two hard rungs attached to the
photon line, which have to be calculated exactly.  The lower part of
the diagram can be then expressed through the gluon distribution in
the target since the asymmetry between the gluons becomes unimportant
in the softer blob. The proof is the same as for the vector meson
electroproduction.
\footnote{ We are indebted to A.Mueller  for the discussion of
this problem.} The cross section is proportional to
\begin{eqnarray}
& &\left. {d\sigma^{\gamma^*+N \rightarrow jet_1 + jet_2 +N}\over dt}
\right| _{t=0}   \propto \ \ \ \ \ \ \ \ \ \  \nonumber \\
& & \left|A_{\gamma^*+gg \rightarrow jet_1+jet_2}\right|^2
\left| \tilde x  G_N(\tilde x, Q^2)\right|^2  \nonumber \\
& & \propto \left( { \alpha_s(Q^2) \tilde x  G_N(\tilde x, Q^2) \over Q^2}
\right)^2  \ ,
\label{eq:35c}
\end{eqnarray}
where $\tilde x$ is the average $x$ of the gluons in the $\gamma^*+gg
\rightarrow jet_1+jet_2$ amplitude, $\tilde x\gg x$, and
$A_{\gamma^*+gg \rightarrow jet_1+jet_2}$ is the hard scattering
amplitude (which includes
at least
2 hard rungs) calculated in PQCD.

 One of the nontrivial predictions of QCD is that the decomposition of
 the cross section for a longitudinally polarized photon in powers of
 $Q^2$ becomes inefficient at small $x$. This is because additional
 powers of $1/Q^2$ are compensated to a large extent by the increase
 with $Q^2$ of $\left[\alpha_s(Q^2)xG(x,Q^2)\right]^2 \sim {Q\over x}$
(see  equations~(\ref{eq:35c}), (\ref{eq:34d1})).  Thus
the prediction of QCD is that electroproduction of hadron states with
$M_{X}^2\ll Q^2$ by longitudinally polarized photons, formally a higher
twist effect, should in practice depend on $Q^2$ rather mildly.  The
contribution of such higher twist effects to the total cross section
for diffractive processes may be considerable, as high as $ 30-40 \%
$.  One of the observed channels, the electroproduction of $\rho$
mesons, constitutes probably up to $10 \% $ of the total cross section
for diffractive processes. So far a detailed quantitative analysis of
this important issue is missing. On the experimental side, it would be
extremely important to separate the longitudinal and transverse
contributions to diffraction.
\section{Can diffractive cross sections raise forever?}
We have demonstrated above that cross sections of hard diffractive processes
are related to cross section of interaction of small color dipole with the
target which
increase fast with incident energy. However such
increase cannot be sustained forever. The simplest way to obtain an upper
limit for
the range of energies where such increase should stop we  consider here the
scattering
of a small object,
a $q\bar q$ pair, from a large object, a nucleon. If only hard physics
was  relevant for the increase of parton distributions at small $x$,
the radius of a nucleon should not increase (small Gribov diffusion).
Under this assumption the unitarity limit corresponds to a black
nucleon.  In this case the inelastic cross section cannot exceed the
geometrical size of the nucleon
 \begin{equation}
\sigma(q\bar q N)= \frac{\pi^2}{3} b^2\alpha_s(1/b^2)
xG_N(x,b^2) < \pi r_N^2 \ .
\label{60d}
\end{equation}
To find the value of $r_N$ in eq.(\ref{60d}) we
 use the optical
theorem to calculate the elastic cross section for a $q\bar q$ pair
scattering off a nucleon,
\begin{equation}
\sigma_{el}={\sigma_{tot}^2\over 16\pi B}
\label{61d}
\end{equation}
where $B$ is the slope of the elastic amplitude (cf. discussion in
section 4). It follows from Eqs.(\ref{60d}),(\ref{61d}) and condition that
$\sigma_{inel} + \sigma_{el}=\sigma_{tot}$ that
the unitarity limit is achieved when
the elastic cross section is equal to the inelastic cross section
$\sigma_{el} \le \sigma_{inel}$. Based on this we find $r^2_N=4 B
\simeq 16{\rm~GeV}^{-2} \simeq (0.8 {\rm~fm})^2$ is the radius of a nucleon.
It follows from the above equations that practically the same estimate
is obtained from the assumption that ${\sigma(el)\over\sigma(tot)}\sim
(0.3-0.5)$.

Applying these inequality for the cases of $\sigma_L$ and $\rho$-meson
production we
find \cite{FKS} that unitarity limit is
reached for  $x_{\sigma_L}(Q^2=5 GeV^2) \sim 3~10^{-5}$,
 $x_{\sigma_L}(Q^2=10 GeV^2) \sim 6~10^{-6}$,  $x_{\rho}(Q^2=5 GeV^2) \sim
3~10^{-4}$,
$x_{\rho}(Q^2=10 GeV^2) \sim 2~10^{-4}$.

The use of the amplitude for $q\bar q$ pair scattering off a nucleon
to deduce the limit allows to account accurately for
nonperturbative QCD effects through the unitarity condition for such
an amplitude.  On the other hand if the increase of parton
distributions is related to soft physics as well then the cross
section may be allowed to increase up to smaller $x$ values.

The black disc limit for $\sigma_{\gamma^*N}$ has been discussed
earlier (for a review and references see~\cite{BKCK,Levin}). The
difference compared to previous attempts is that we deduce the QCD
formulae for the cross section of a $q\bar q$ pair scattering off a
hadron target.  For this cross section the geometrical limit including
numerical coefficients unambiguously follows from unitarity
 of the $S$-matrix,
that is
the geometry of the collision. As a result we obtain an inequality
which contains no free parameters.  Recently a quantitative estimate
of the saturation limit was obtained~\cite{Collins} by considering the
GLR model~\cite{LR,MQ} of nonlinear effects in the parton evolution
and requiring that the nonlinear term should be smaller than the
linear term.  The constraint obtained for $xG_p(x,Q^2)$ is numerically
much less restrictive compared  to our result.
Even a more stringent restriction follows
for the interaction of a colorless gluon pair off a nucleon from the
requirement that the inelastic cross section for the scattering of a
small size gluon pair should not exceed the elastic one
\begin{equation}
\sigma(gg N)= \frac{3\pi^2}{4} b^2\alpha_s(1/b^2)
xG_N(x,b^2) <\pi r_N^2 \ .
\label{60f}
\end{equation}
For $b=0.25{\rm~fm}$ the geometrical limit is achieved for $x\sim
10^{-3}$.

We want to point out that the black disc limit implies a restriction
on the limiting behavior of the cross sections for hard processes but
does not allow to calculate it. The dynamical mechanism responsible
for slowing down of the increase of parton distributions so that they
satisfy equations~(\ref{60d},~\ref{60f}) is not clear.  In particular
the triple Pomeron mechanism for shadowing suggested in~\cite{LR} does
not lead to large effects at HERA energies especially if one assumes a
homogeneous transverse density of gluons~\cite{Levin,Martin}.

The theoretical analysis performed in this section does not allow to
deduce restrictions on the limiting behavior of parton distributions
in a hadron. Beyond the evolution equation approximation and/or
leading $\alpha_s \ln x \ln {Q^2\ over \Lambda_{QCD}^2}$ approximation
the restriction on the cross
sections of deep inelastic processes cannot be simply expressed in
terms of parton distributions in a hadron target.

We want to draw attention to the fact that nonperturbative QCD effects
play an important role in the contribution of higher twist effects to
$\sigma_L(\gamma^* p)$.  This is evident from the impact parameter
representation of the contribution to $\sigma_L(\gamma^* +p)$ of $n$
consecutive rescatterings of small transverse size $q\bar q$
pairs.  This contribution is proportional to
$$Q^2\int |\psi_{\gamma_L}^*(z,b^2)|^2 dz  d^2b
\left[\alpha_s(1/b^2)b^2xG(x,b)\right]^{n} \ .$$ The
  inspection of this integral shows that for large $n\geq 3$, $b$ which
  dominates under the integral does not decrease with increasing $Q^2$
for $x \sim 10^{-3} \div 10^{-4}$.
We use as estimate $\alpha_sxG_N(x,Q^2) \propto \sqrt{Q} $ which follows
from the evolution equation for small $x$.
(This QCD effect is absent in the applications~\cite{NZZ} of the
constituent quark model).
Thus if higher twist effects were really important in small $x$
physics, it would imply that the small $x$ physics is the
outcome of an interplay of hard (small $b$) and soft (large $b$) QCD.
To illustrate this point let us consider the cross section of
diffractive electroproduction of hadrons with masses $M^2\sim Q^2$ by
transversely polarized photons. Applying the same ideas as in the case
of longitudinally polarized photons we would obtain a similar
expression as given by equation~\ref{eq:34d1}. The important
difference is that the wave function of a transversely polarized
photon is singular for $z\rightarrow 0$ or $1$. As a result the
contribution of large impact parameters $b$ in the wave function of
the photon should give the dominant contribution to the integral in a
wide kinematical range of $x$ and $Q^2$. This has been understood long
ago -- see discussion in sections 8-9.  A similar conclusion has
been achieved recently~\cite{NZZ} within the constituent quark model.
(Note however that this model ignores the increase of gluon
distribution with $Q$ typical for QCD and therefore overestimates the
nonperturbative QCD contribution). Thus such type of diffractive
processes should depend on energy in a way similar to the usual soft
hadron processes.

A good example of the consequences of the interplay of small $b$ and
large $b$ physics is that in electroproduction of small mass states
the unitarity limit may become apparent at larger $x$
than in the case of the total cross section of deep inelastic processes.

\section{Diffraction in DIS at intermediate $Q^2$}

It has been understood long ago that the production of almost on mass
shell quarks by virtual photons should give a significant contribution
to the total cross section for deep inelastic scattering at small
$x$~\cite{BJ71}.  One of the predictions of this approach (which is
essentially the parton model approximation) is a large cross section
for diffractive processes. The  QCD $Q^2$ evolution does not change this
physical picture radically. The only  expected modification of the picture is
the appearance   of a number of hard jets in
the current fragmentation region~\cite{FS88} typical for
including  $\alpha_S \ln Q^2$ terms.  It is often stated   that the
dominance of the BFKL Pomeron in diffractive processes
predicts the dominance of final states consisting
of hard jets ~\cite{Ryskin2,LW}.  However this prediction is not robust since
 the analysis of Feynman diagrams for hard processes
in QCD finds strong diffusion effects into the region of small
transverse momenta of partons (see~\cite{Bartels} and references
therein).  Recent HERA data~\cite{MD} seem to support the picture with
a dominance of events with small $k_t$. Thus it seems worthwhile to
investigate the role of nonperturbative QCD physics in diffractive
processes.

The interaction of a virtual photon with a target at intermediate
$Q^2$ and small $x$, when gluon radiation is negligible, can be
considered as a transformation of $\gamma^*$ into a $q \bar q $ pair
which subsequently interacts with the target.  In this case an
important role is played by the quark configurations in which the
virtuality of the quark interacting with the target is small,
\begin{equation}
k_{qt}\sim  k_{t0} \ ,\ \ \ \alpha_q = {(m_q^2 +k_{qt}^2)\over Q^2} \ .
\label{soft}
\end{equation}
Here $\alpha_q$ denotes the light-cone fraction of the photon momentum
carried by the slower quark and $k_{t0}$ is an average transverse
momentum of partons in the hadron wave function, typically $k_{t0}
\sim 0.3-0.4 $~GeV.

In the language of non-covariant perturbation theory the $q\bar q$
configurations described by~(\ref{soft}) correspond to an intermediate
state of mass $m^2 \sim Q^2$ and of transverse size $\sim {1 \over
  k_{t0} } \geq 0.5$ fm.  These configurations constitute a tiny
fraction $\sim {k_{qt}^2\over Q^2}$ of the phase volume kinematically
allowed for the $q \bar q$ pair.  However the interaction in this case
is strong -- similar to the interaction of ordinary hadrons, since the
virtuality of the slower quark is small and the transverse area
occupied by color is large.  The contribution of these configurations
leads to Bjorken scaling since the total cross section is proportional
to ${1\over Q^2}$ and in the parton model only these configurations
may contribute to the cross section. Hence Bjorken has
assumed~\cite{BJ71} that all other configurations are not important in
the interaction though the underlying dynamics of such a suppression
was not clear at that time~\cite{BjKogut}.  Accounting for the $
({k_{qt}^2\over Q^2})$ factor in  the Gribov dispersion representation
 allowed him
to reconcile this  dispersion representation with scaling.  He
suggested to refer to these configurations as aligned jets since both
quarks have small transverse momenta relative to the photon momentum
direction. In further discussions we will refer to this approach as
that of the Aligned Jet Model (AJM).  Note that in terms of the
Feynman fusion diagram the aligned jet contribution arises only for
transversely polarized virtual photons. This is because the vertex for
the transition $\gamma^*_T \rightarrow q \bar q$ is singular $ \sim
{1\over z}$ when the fraction of the photon momentum $z$ carried by
the slowest quark (antiquark) tends to 0. For the case of a
longitudinally polarized photon the
naive
aligned jet approximation produces
results qualitatively different from expectations in QCD where the
contribution of symmetric jets dominates. This is because in QCD the
dominant contribution to the $\gamma^*_L$--nucleon cross section
arises from the region of large $k_{qt} \sim {Q\over 2}$.

In QCD the interaction of quarks with large relative transverse
momenta with a target is suppressed but not negligible. The
suppression mechanism is due to color screening since $q \bar q$
configurations with large $k_t$ correspond, in the coordinate space,
to configurations of small transverse size, $b \sim {1\over k_t}$, for
which equation~(\ref{eq:9c}) is applicable. It is easy to check that
the contribution of large $k_t$ also gives a scaling contribution to
the cross section. The practical question then is which of the two
contributions dominates at intermediate $Q^2 = Q_0^2 \approx 4
$~GeV$^2$, above which one can use the QCD evolution equations. To
make a numerical estimate we assume that the $q \bar q$ configurations
with $k_{qt} \leq k_{t0}$, in which color is distributed over a
transverse area similar to the one occupied by color in mesons,
interact with a cross section similar to that of the pion. A
comparison with experimental data for $F_{2p}(x \sim 0.01, Q^2_0) $
indicates that at least half of the cross section is due to soft, low
$k_t$ interactions~\cite{FS88,FS89}.  A crucial check is provided by
applying the same reasoning to scattering off nuclei in which the
interaction of the soft component should be shadowed with an intensity
comparable to that of pion-nucleus interaction.  Indeed the current
deep inelastic data on shadowing for $F_{2A}(x,Q^2)$ are in
reasonable agreement with calculations based on the soft mechanism of
nuclear shadowing~\cite{FS88,FS91}.

Similarly to the case of hadron-nucleon and hadron-nucleus
interactions, the interaction of $\gamma^*$ in a soft hadron component
naturally leads to diffractive phenomena. Application of the Gribov
representation with a cutoff on the $k_t$ of the aligned jets in the
integral leads to a diffractive mass spectrum for the transversely
polarized virtual photon~\cite{BjKogut}
\begin{equation}
{d \sigma \over d M^2} \propto {1\over (M^2+Q^2)^2} \ .
\label{dajm}
\end{equation}
The two major differences compared to the hadronic case are that
elastic scattering is substituted by production of states with $\left<
M^2\right> \approx Q^2$
and that the contribution of configurations of small spatial size
is larger for $\gamma^*_L$.

If the aligned jet configurations were dominant, the fraction of cross
section of deep inelastic $\gamma^*N$ scattering due to single
diffractive processes would be
\begin{equation}
R_{\rm single~dif}^{\rm AJM}= {\sigma_{\rm dif}\over \sigma_{\rm tot}}=
  {\sigma_{\pi N}({\rm el})+\sigma_{\pi N}({\rm dif})\over
\sigma_{\pi N}({\rm tot})} \sim 0.25. \ \
\label{eq:43c1}
\end{equation}
Our numerical estimates indicate that for $Q^2 \sim Q^2_0$ and $ x
\sim 10^{-2}$ the AJM contributes about $\eta \sim 60-70 \%$ of the
total cross section. So we expect that in this $Q^2$ range the
probability for diffraction is
\begin{equation}
R_{\rm single~dif}= \eta R_{\rm single~dif}^{\rm AJM} \sim 15\%.
\end{equation}
This probability is actually related in a rather direct way to the
amount of shadowing in interactions with nuclei in the same kinematic
regime, so it is quite well determined by the nuclear shadowing data.

To estimate the probability of events with large rapidity gaps one has
to add the processes of diffractive dissociation of the nucleon and
double diffraction dissociation, leading to an estimate
\begin{equation}
P_{\rm gap} =(1.3-1.5) R_{\rm single~diff} \sim 0.2 \ .
\label{44}
\end{equation}
This is rather close to the observed gap survival probability for
photoproduction processes~\cite{photogap}.

The characteristic features of the AJM contribution which can be
checked experimentally are the charge and flavor correlations between
the fastest and the slowest diffractively produced hadrons which
should be similar to those in $e^+e^-\rightarrow$~hadrons at $M^2\sim
Q^2$.

Another important feature of the soft contribution which distinguishes
it from the contribution of hard processes is the $t$ dependence of
the cross section for $M^2 \le Q^2$. Since the size of the
configurations is comparable to that of the pion one may expect that
the $t$ slope of the cross section, $B$, should be similar to that of
the pion-nucleon interaction, i.e. $B \ge 10$~GeV$^{-2}$ which is
much softer than for hard processes where we expect $B \approx
4$~GeV$^{-2}$ (see discussion in section 3).  The large value of
the slope for the soft component is also natural in the parton type
logic where only slow partons interact. It is easy to check that for
$-t \gg k_{t0}^2 \sim 0.1$~GeV$^2$ the mass of the produced hadron
system is larger than the mass of the intermediate state by factor
${\sqrt{-t} \over k_{t0}}$. Thus for large $t$ the production of
masses $M \le Q$ is suppressed. Therefore the study of the
$t$-dependence of diffraction can be used to disentangle the
contribution of soft and hard mechanisms.

This discussion indicates also that the contribution of non-diagonal
transitions $"M^2" \rightarrow "M^{\prime~2}"$ leads to a weaker decrease
of the differential cross section with $M^2$ than given by
equation~(\ref{dajm}). Besides at large $M^2 \sim ~few ~Q^2$ one expects
an onset of the dominance of the triple Pomeron mechanism which corresponds
to
\begin{equation}
{d \sigma \over d M^2} \propto {1 \over Q^2 M^2}.
\label{PPP}
\end{equation}
\section{$Q^2$ evolution of the soft contribution
in diffraction.}

The major difference between the parton model and QCD is the existence
in QCD of a significant high $p_t$ tail in the parton wave functions
of the virtual photon and the proton. This is the source of the
violation of Bjorken scaling observed at small $x$.  It is thus
necessary to modify the aligned jet model to account for the hard QCD
physics.

It is in general difficult to obtain with significant probability a
rapidity gap in hard processes in perturbative physics.  Confinement
of quarks and gluons means that a gap in rapidity is filled by gluon
radiation in PQCD and subsequently by hadrons~\cite{Feynman71}.  It is
possible to produce diffraction in perturbative QCD but the price is a
suppression by powers of the coupling constant $\alpha_s$ and/or
powers of $Q^2$.  In first approximation in calculating diffraction in
deep inelastic processes at small $x$ we will thus neglect diffraction
in PQCD. In the following analysis, for the description of large
rapidity gap events, we shall use the QCD modification of the AJM
model suggested in~\cite{FS88} as well as the suggestion of
Yu.Dokshitzer~\cite{Doc94} to add to the conventional evolution equation
the assumption of local duality in rapidity space between quark-gluon
and hadron degrees of freedom.

In the course of the following considerations it will be convenient to
switch to the Breit frame.  In this frame the photon has momentum
$(0,-2xP)$ and the initial proton has momentum ($P$,$P$).
Correspondingly $Q^2=4x^2P^2$. The process of diffraction can be
viewed as the virtual photon scattering off a color singlet $q \bar q$
pair with the interacting parton carrying a light-cone fraction
$\alpha$ and the spectator parton carrying a light-cone fraction
$x_1$. We assume here the local correspondence in rapidity space
between partons and hadrons. The mass of the produced system, $M$ is
given by
\begin{eqnarray}
 M^2 & = & (p_{\gamma^*}+p_{x_1}+p_{\alpha})^2=
%P^2((\alpha+{x_1} )^2-(\alpha+{x_1}-2x )^2)=
\nonumber \\
& & Q^2+4P^2 (\alpha + x_1)x= Q^2 {\alpha + x_1 -x \over x} \ .
\label{B1}
\end{eqnarray}
In the approximation that gluon radiation is neglected (parton model)
$\alpha=x$ and the mass of the diffractively produced system $M$ is
\begin{equation}
M^2=Q^2 x_1/x \ .
 \label{B2}
\end{equation}
The differential cross section for production of mass $M$ follows from
equation~(\ref{dajm}),
\begin{eqnarray}
{d \sigma^{\rm AJM}\over d  M^2} =
\Gamma \int d x_1 \delta (x_1 -{x M^2\over Q^2}){1 \over  (Q^2
+M^2)^2}= \nonumber \\
{\Gamma
\over Q^4} \int d x_1 \delta(x - \alpha)
\delta
 (x_1 - {x M^2 \over Q^2})
{1 \over (1 + x_1/x)^2} \ .
\label{B3}
\end{eqnarray}
Here $\Gamma$ is the factor which includes the density of correlated
color singlet pairs and the cross section for interaction of the
photon with the parton.  The total cross section for diffractive
dissociation comes out to be proportional to ${1 \over Q^2}$,
\begin{equation}
\int {d\sigma^{AJM} \over dM^2}
 d M^2={\Gamma \over Q^2}\int
 {d x_1 \over x}
{1 \over (1 + x_1/x)^2} = {\Gamma
\over
 Q^2} \ .
\label{B4}
\end{equation}
We do not restrict the integration over $x_1$ in equation~(\ref{B4})
since the major contribution comes from the region of $x_1\sim x$.
Thus we can formulate diffraction in the infinite momentum frame as a
manifestation of short rapidity range color correlation between
partons in the nonperturbative parton wave function of the nucleon. To
calculate the $Q^2$ evolution in QCD we have to take into account that
the parton with momentum fraction $\alpha$ has its own structure at
higher $Q^2$ resolution and that the $\gamma^*$ scatters off
constituents of the "parent" parton. This is the usual evolution with
$Q^2$ which can be accounted for in the same way as in the QCD
evolution equations by the substitution
$$\Gamma \delta (x-\alpha) \rightarrow P\sum\limits_{j} e_j^2 d_j^{\rm
  pert}({x \over \alpha},Q^2,Q_0^2)$$ where $d_j^{\rm
  pert}(x,Q^2,Q_o^2)$ are the structure functions of the parent
parton.  This effect leads to the change of the relationship between
$x_1$ and $x$ resulting from parton bremsstrahlung.
After performing the integral over $x_1$ we obtain
\begin{eqnarray}
& & {d \sigma^{\rm soft+QCD}
\over d M^2} =\nonumber  \\
& & {P \over Q^4}
\int^1_x
{d\alpha \over \alpha}  \sum\limits_{j}  e^2_j
d_j^{\rm pert}({x \over \alpha},Q^2,Q^2_0)
d_j^{\rm nonpert}(\alpha, Q_0^2)\nonumber  \\
& & {1 \over ( 2-\alpha/ x+M^2/Q^2 )^2}\times\nonumber  \\
 & & \theta \left( \lambda (1-x) -{xM^2 \over Q^2} \right) \theta
\left( 1-\frac{\alpha}{x} +\frac{M^2}{Q^2}\right) \ .
\label{B7}
\end{eqnarray}
Here $P$ denotes the probability of diffractive scattering in a soft
interaction and $d_j^{\rm nonpert}(\alpha, Q_0^2)$ is the parton
distribution in the soft component producing diffraction (compare
discussion in the previous section).  The $\theta$ function term
reflects the condition that diffraction in the nonperturbative domain
is possible only for
\begin{equation}
0 \leq {M^2 \over s}
={\alpha +x_1-x \over (1-x)}\equiv   \lambda
\sim 0.05-0.1 \ \ .
\label{B6}
\end{equation}

\subsection{Qualitative pattern of $x$ and  $Q^2$ dependence of diffraction.}

It is easy to see that discussed  equations  lead to
the leading twist diffraction.  To see the pattern of the $x,Q^2$
dependence we can assume that $d_j^{\rm pert}(x,Q^2)={d\over x^n}$ and
$d_j^{\rm nonpert} (x,Q_0^2)={d_0\over x^{n_0}}$.  It follows from
equation~(\ref{B7}) that for $x\ll \lambda $ the ratio ${\sigma_{\rm
    diff} \over \sigma_{\rm tot}}$ does not depend on $x$. One can
also see that the characteristic gap interval is
\begin{eqnarray}
\Delta  y= \ln\frac{s}{Mm_p}==   \ln {1\over x} +
\ln({Q^2\over M~m_p}) \ .
\label{B9}
\end{eqnarray}
The second term $\ln\frac{Q^2}{M~m_p}$ increases with $Q^2$ in the
parton model, while the scaling violation tends to reduce this
increase since the mean value of $M^2/Q^2$ at fixed $x$ increases with
$Q^2$.

There are several qualitative differences between the QCD improved
soft diffraction and the parton model (AJM).

\noindent
(i) Due to QCD evolution the number of diffractively produced hard
jets and the average transverse momentum of diffractively produced
hadrons should increase with $Q^2$.

\noindent
(ii) The distribution of ${M^2\over Q^2}$ becomes broader in QCD with
increasing $Q^2$.

\noindent
(iii) While in the parton model the cross section for the interaction
of the longitudinally polarized virtual photon is a higher twist
effect, in QCD diffraction is a leading twist for any polarization of
the virtual photon. The final state in the case of longitudinally
polarized photons should contain at least 3 jets, two of them should
have large transverse momenta comparable with $Q$.

\noindent
(iv) The $x$ dependence of the soft component is likely to be faster than
for soft Pomeron as seen in $pp$ scattering both due to smaller
screening corrections and due to contribution of configurations
of sizes somewhat smaller than  normal hadron sizes. In the laboratory
frame of the target  these configurations correspond to
$q \bar q$ pairs with $p_t \sim 0.5 \div 1 GeV/c$.

\subsection{Connection with the Ingelman-Schlein Model}
Ingelman and Schlein have suggested to treat hard diffractive
processes using the concept of parton distribution in the Pomeron
\cite{IS}.  In this approach one calculates the light-cone fraction of
the target carried by the Pomeron, $x_P$, and light-cone fractions of
the Pomeron momentum carried by quarks and gluons, $\beta$. It is
assumed that parton distributions in the Pomeron, $\beta q_P(\beta,
Q^2), \beta g_P(\beta, Q^2)$ are independent of $x_P$ and the transverse
momentum of the recoil nucleon. For the process of inclusive deep
inelastic diffraction $\beta $ is simply related to the observables,
\begin{equation}
\beta={Q^2 \over Q^2 + M_X^2}
\label{beta}
\end{equation}
The $Q^2$ evolution of the total cross section of diffraction as
considered in the previous subsections is consistent with the
expectation of the Ingelman-Schlein model (though the final states are
not necessarily the same).  The aligned jet model in this case serves
as a boundary condition defining parton distributions in the Pomeron
at intermediate $Q_0^2$ above which QCD evolution takes place.  The
aligned jet model corresponds to the quark distribution in the Pomeron
\begin{equation}
 \beta q_P(\beta, Q_0^2) \propto \beta .
\label{qbeta}
\end{equation}
It follows from the discussion in the end of section 8 that taking
into account the non-diagonal transitions in the aligned jet model and
the triple Pomeron contribution would make the distribution flatter.
A similar, rather flat, distribution is expected for gluons for these
$Q^2$.  This expectation of the aligned jet model is different from
the counting rule anzatz of \cite{IS}:
$ \beta q_P(\beta, Q_0^2) \propto (1-\beta)$.  Note also
that since the density of gluons at small $x$ is larger than
the $q \bar q$ density
 the effective gluon density in the Pomeron should be larger
 that the charged parton
density already at $Q^2 \sim Q^2_0$.

\section{Non-universality of the  pomeron in QCD.}

Theoretical considerations of soft diffractive processes have
demonstrated that ordinary hadrons contain components of very
different interaction strength~\cite{BBFHS93,BBFS93}.  This includes
configurations which interact with cross sections much larger than the
average one and configurations which interact with very small cross
sections, described by equation~(\ref{eq:9c}) for a meson projectile.

The presence in hadrons of various configurations of partons having
different interaction cross sections with a target is in evident
contradiction with the idea of a universal vacuum pole where universal
factorization is expected. At the same time it is well known that the
Pomeron pole approximation is not self-consistent. The vacuum pole
should be accompanied by a set of Pomeron cuts~\cite{Gribov4}.  For
the sum of the Pomeron pole and the Pomeron cuts no factorization is
expected.  Thus the S matrix description and the QCD description are
not in variance.  We shall enumerate now where and how to search for
the non universality of the effective Pomeron understood as the sum of
the pomeron pole and the Pomeron cuts.

It is natural to distinguish two basic manifestations of the
non universality of the effective Pomeron trajectory, $\alpha_{\cal
  P}(t) \approx \alpha_0 + \alpha't$, a different energy dependence of
the interaction cross section, which is characterized by a different
value of $\alpha_0$, and a different rate of the Gribov diffusion,
which would manifest itself in different values of $\alpha'$.

\subsection{Non-universality of the energy dependence.}

To study the non universality of $\alpha_0 $ it is necessary to study
the energy dependence of the electroproduction of vector mesons as a
function of $Q^2$. Up to now only two results are known, $\alpha_0
\sim 1.08$ from the $\rho$ meson photoproduction~\cite{photorho}, and
$\alpha_0 \sim 1.30$ as estimated from  NMC and  HERA data at $Q^2
\sim 10$~GeV$^2$~\cite{NMC1,ZEUSb}.  The key question is at what $Q^2$
a significant rise of $\alpha_0 $ starts -- this will give a direct
information on the transition region from soft to hard physics.
A fast increase of $F_{2p}(x,Q^2)$ at small $x$
observed
at $Q^2$ as low as 1.5 GeV$^2$ indicates  that the rise may occur
already at $Q^2 \sim
3$~GeV$^2$. The same question applies for production of heavier
$\phi$ and $J/\Psi$ mesons. Since the $J/\Psi$ meson is a small object
one may speculate that in this case the rise could start already for
photoproduction (the experimental data indicate that the slope of the
$J/\Psi$ exclusive photoproduction cross section is close to the value
given by the two-gluon form factor of the nucleon). The practical
problem for a quantitative analysis is that no accurate
data on {\bf exclusive} $J/\Psi$ photoproduction at fixed target
energies are available at the moment. Inclusive fixed target data
where the $J/\psi$ meson carries practically the whole momentum of the
projectile photon which are used to extract the exclusive channel seem
to be significantly contaminated by the contribution of the reaction
$\gamma + p \rightarrow J/\Psi + X$ which is peaked at $x_F \equiv
p_{J/\Psi}/p_{\gamma}$ close to 1.

\subsection{Non-universality of the t-dependence.}

The slope of the effective Pomeron trajectory $\alpha'$ should
decrease with increasing $Q^2$. This is because the Gribov diffusion
in the impact parameter space, which leads to finite
$\alpha'$~\cite{Gribovdif}, becomes inessential in the hard regime.
This is a consequence of the increase with energy of the typical
transverse momenta of partons. Thus for the reactions $\gamma^* +
N\rightarrow V + N$ the effective $\alpha'$ should decrease with
increasing $Q^2$ while a universal Pomeron exchange approximation
predicts for the energy dependence of the slope
\begin{equation}
B(s)=B(s_0) + 2 \alpha' \ln \left( {s\over s_0} \right)
\end{equation}
with $\alpha' \sim 0.25$~GeV$^{-2}$.
It is possible to look for this effect by comparing the HERA and the
NMC data on the $\rho$ meson production. The universal Pomeron model
predicts that the slope should change by $\Delta B \sim 2 ~GeV^{-2}$
between NMC energies where   $B \sim
4\div 5$~GeV$^{-2}$~\cite{NMC1} and  HERA
energies while in the perturbative domain a much weaker change of the
slope is expected.

The slope of the effective Pomeron trajectory $\alpha'$ may depend on
the flavor. It should decrease with the mass of flavor.  Thus it would
be very important to measure the effective $\alpha'$ for the
diffractive photoproduction of $\rho, ~\phi$ and $J/\Psi$.  If PQCD is
important for $J/\Psi$ photoproduction one would expect a smaller
increase of the slope with energy in this case.

\subsection{Non-universality of the gap survival probability.}

The presence of configurations of different size in hadrons (photons)
should also manifest itself in the non universality of the gap survival
probability in the two jet events.  Since the probability of gap
survival is determined by the intensity of the {\it soft} interaction
of the projectile with the target, the survival probability should
increase with increase of $Q^2$, and at fixed $Q^2$ it should be
larger for the heavy $q \bar q$ components of the photon.  Also, the
gap survival probability in the photon case should be substantially
larger than that observed in $p \bar p$ collisions at
FNAL collider~\cite{gapfnal}. This reflects the difference between
$\sigma_{tot}(p \bar p) \approx 80$ mb and the effective cross section
for the interaction of the hadronic components of $\gamma (\gamma^*)$
with nucleon of $\le 30$ mb.

Observation of non-universalities discussed here will shed light on the
structure of the effective Pomeron operating in strong interactions
and will help to address the question about the major source of the
increase of the total cross section of $p \bar p$ interaction --- soft
physics or hard physics of small size configurations.

\subsection{Non-universality of diffraction dissociation}

Since the object which couples to the nucleon in the hard coherent processes
is different from soft Pomeron one may expect a difference between  the value
of the
ratio $\left. {{d \sigma^{\gamma^* +p \rightarrow \rho +X} \over dt} \over
{d \sigma^{\gamma^* +p \rightarrow \rho +p} \over dt}}\right|_{t=0}$ and
similar ratio
for soft processes. Qualitatively, one may expect that since the coupling of
effective Pomeron in hard processes
is more local the ratio of diffraction dissociation and elastic cross sections
should  be substantially smaller  for hard  processes, at least for small
excitation masses.

\section{Summary.}

We have demonstrated that color coherent phenomena should play in QCD
a rather direct role both in the properties of hadrons and in the high
energy collisions. It seems now that recent experimental data confirm
some of the rather nontrivial predictions of QCD and help to elucidate
such old problems as the origin of the Pomeron pole and the Pomeron
cuts in the Reggeon Calculus. Thus we expect that the investigation of
coherent hard and soft diffractive processes may be the key in
obtaining a three dimensional image of hadrons, in helping to search
for new forms of hadron matter at accelerators and in understanding
the problem of inter-nucleon forces in nuclei. Forthcoming high
luminosity studies of diffraction at HERA which will include among
other things the detection of the diffracting nucleon and the $\sigma_L -
\sigma_T$ separation would greatly help in these studies.

 \section*{Acknowledgments}

 We would like to thank H.Abramowicz, J. Bartels, J. Bjorken, S. Brodsky, W.
 Buchmuller, A. Caldwell, J. Collins, J. Ellis, G. Knees, H. Kowalski,
 L. Lipatov, A. Mueller, G. Wolf for the fruitful discussions of the
 interplay of soft and hard physics and of methods of their
 investigation.


\vspace{1cm}
\Bibliography{100}
\bibitem{AFS} H.Abramowicz,  L. Frankfurt and M. Strikman, DESY-95-047;
Proceedings of the SLAC 94 summer school, in press.
\bibitem{FMS94} L. L. Frankfurt, G. A. Miller and M. Strikman, Ann. Rev.
of Nucl.
and Particle Phys.~44~(1994)~501.
\bibitem{BBFS93}B. Bl\"{a}ttel, G. Baym, L. L. Frankfurt and M. Strikman,
Phys. Rev. Lett.~71~(1993)~896.
\bibitem{FMS93}L. Frankfurt, G. A. Miller and M. Strikman, Phys. Lett.
B304~(1993)~1.
\bibitem{FKS} L.Frankfurt, W.Koepf, and M.Strikman,
Preprint TAUP-2290-95, hep-ph/9509311.
\bibitem{Low} F. E. Low, Phys. Rev. D12~(1975)~163.
\bibitem{Nussinov} S. Nussinov, Phys. Rev. Lett. 34~(1975)~1286.
\bibitem{GS} J. Gunion and D. Soper, Phys. Rev. D15~(1977)~2617.
\bibitem{Brod94}S. J. Brodsky, L. Frankfurt, J. F. Gunion, A. H. Mueller and
M. Strikman, Phys. Rev. D50~(1994)~3134.
\bibitem{NMC1}NMC Collaboration,
M. Arneodo et al.,Nucl.Phys. B429 (1994)~502
\bibitem{ZEUSb} M.Derrick et al., DESY-95-133, July 1995
\bibitem{DL1} A.Donnachie and P. V. Landshoff,
 Phys. Lett.\\~185B~(1987)~403; Nucl. Phys. B311~(1989)~509.
\bibitem {psi1} P. L. Frabetti et al., Phys. Lett. B316~(1993)~197;
 M. Binkley et al., Phys. Rev. Lett. 48~(1982)~73.
\bibitem{Asratian}  A. E. Asratian et al., Z. Phys. C58~(1993)~55.
\bibitem{Gribovdif}V. N. Gribov, %in ''Regge theory of
%low-p(T) hadronic interactions'', Caneschi L. (ed.) 1990, p. 22-23;
JETP Lett. 41~(1961)~667;
Yad. Fiz. 5~(1967)~399;  Yad. Fiz. 9~(1969)~3;
'' Space-time  description  of hadron   interactions  at high energies'',
1st ITEP school, v.I ''Elementary particles'', p. 65 (1973).

\bibitem{Ryskin} M. G. Ryskin, Z. Phys. C37~(1993)~89.
\bibitem{F94} L. Frankfurt, invited talk
at the Minischool on Diffractive Physics, DESY, May 1994.
\bibitem{FS89} L. L. Frankfurt and M. Strikman, Phys. Rev. Lett.
64~(1989)~1914.
\bibitem{MuTan} A. H. Mueller and W. K. Tang, Phys. Lett. B284~(1992)~123
\bibitem{FR}  J. R. Forshaw and M. G. Ryskin, hep-ph/950376.
\bibitem{RPP} Review of Particle Properties, Phys. Rev. D50~(1994)~1177.
\bibitem{GLDAP}
V. N. Gribov, L. N. Lipatov,  Yad. Fiz. 15~(1972)~781; Yad. Fiz. 15~(1972)
{}~1213;
Yu.L. Dokshitzer,  Sov. Phys. JETP 46~(1977)~641;
G. Altarelli and G. Parisi, Nucl. Phys. B126~(1977)~298.
\bibitem{MR} M. G. Ryskin, Yad. Fiz. 50~(1989)~1428.
\bibitem{CFS93} J. C. Collins, L. L. Frankfurt and M. Strikman, Phys.
Lett B307~(1993)~161.
\bibitem{F92} L. L. Frankfurt, '' Hard diffractive processes at colliders'',
talk at the FAD meeting at Dallas, TX, March 1992.
\bibitem{SB}  A. Berera and D. Soper, Phys. Rev. D50~(1994)~4328.
\bibitem{IS} G. Ingelman and P. Schlein, Phys. Lett. B152~(1985)~256.
%\bibitem{UA8}
\bibitem{Br} UA8 Collaboration, A. Brandt et al., Phys. Lett. B297~(1992)~417.
\bibitem{DL2} A.Donnachie and P.V.Landshoff, Phys. Lett. \\B285~(1992)~172.
\bibitem{XY} M. Diehl, Z.Phys.C66~(1995)~181.
\bibitem{CTEQ} J. C. Collins, J. Huston, J. Pumplin, H. Weerts and J.
  J. Whitmore, Phys.Rev.D51~(1995)~3182.

\bibitem{Ryskin2} M. Ryskin and M. Besancon, in Proceedings of the HERA
Workshop, ''Physics at HERA'', vol.1, edited by W. Buchmuller and G. Ingelman,
  ~(1991)~215.
\bibitem{NZ} N. N. Nikolaev and B. G. Zakharov,  Z. Phys.\\ C53~(1992)~331.

\bibitem{NZZ}N. N. Nikolaev and . G. Zakharov, Z.Phys. \\C64~(1993)~631.

\bibitem{BjKogut}J. D. Bjorken and J. B. Kogut,  Phys. Rev. D8~(1973)~1341.
\bibitem{FS88}L. L.~Frankfurt and M.~Strikman, Phys. Rep. \\160~(1988)~235.
\bibitem{Bartels} J. Bartels, H. Lotter, M. Wusthoff,
  DESY-94-245 (1994).
\bibitem{BKCK} B. Badelek, M. Krawczyk, K. Charchula, J.Kwiecinski,
  Rev. Mod. Phys. 64~(1992)~927.
\bibitem{Levin} E. Laenen and E. Levin, , Ann. Rev.
of Nucl.
and Particle Phys.,44~(1994)~199.
\bibitem{Collins} J. Collins and J. Kwiecinski, Nucl. Phys.
  B335~(1990)~89.
\bibitem{LR} L. V. Gribov, E. M. Levin,
  M. G. Ryskin, \\Phys. Rep.~100~(1983)~1.
\bibitem{MQ} A. H. Mueller and J. Qiu,
  Nucl. Phys. B268~(1986)~427
\bibitem{Martin} A. J. Askew, J. Kwiecinski, A. D. Martin, P. J. Sutton,
Phys. Rev. D49~(1994)~4440
\bibitem{BJ71}J. D. Bjorken  in Proceedings of  the International
Symposium on Electron and Photon Interactions at High Energies,
p. 281--297, Cornell (1971).
\bibitem{MD} ZEUS Collaboration, M. Derrick et al., Phys. Lett.
B332~(1994)~228.

\bibitem{LW} E. M. Levin, M. Wusthoff, Phys. Rev. D50~(1994)~4306.
\bibitem{photogap} ZEUS Collaboration, M. Derrick et al.,\\ Z. Phys.
C63~(1994)~391.
\bibitem{FS91}L. L. Frankfurt and M. Strikman, in ''Modern Topics in
Electron Scattering'', Editors B. Frois and I. Sick, 1991, p.762, World
Scientific.
\bibitem{Feynman71} R. P. Feynman, '' Photon-hadron interactions'',
W. A. Benjamin, Inc, Reading, Massachusetts, 1972.
\bibitem{Doc94}  Yu. Dokshitzer, invited talk at the Minischool on Diffractive
Physics, DESY,
May 1994.
 \bibitem{BBFHS93} B. Bl\"attel, G. Baym, L. L. Frankfurt, H. Heiselberg
and M. Strikman, Phys.  Rev. D47~(1993)~2761.
\bibitem{Gribov4} V. N. Gribov, %in ''Regge theory of low-p(T) hadronic \\
%interactions'', Caneschi, L. (ed.) 1990, p 63-71;\\
Sov.~Phys.~JETP~26~(1968)~414; Zh. Eksp.
Teor. Fiz.~53~(1967)~654.
\bibitem{photorho} M.Derrick et al., DESY 95-143.
\bibitem{gapfnal} S. Abachi et al., Phys. Rev. Lett. 72~(1994)~965.
\end{thebibliography}
\end{document}